\documentclass{aastex}
\usepackage{emulateapj5,graphicx}

\begin{document}
\title{
Formation of Dwarf Galaxies during the Cosmic Reionization \\
}
\author{Hajime Susa\altaffilmark{1} 
\vskip 0.2cm
\affil{Department of Physics, Rikkyo University, Nishi-Ikebukuro,
Toshimaku, Japan}
\vskip 0.3cm
Masayuki Umemura\altaffilmark{2}
\vskip 0.2cm
\affil{Center for Computational Physics, University of
  Tsukuba, Japan }
\altaffiltext{1}{susa@rikkyo.ac.jp}
\altaffiltext{2}{umemura@rccp.tsukuba.ac.jp}
}

\begin{abstract}
We reanalyze the photoevaporation problem of subgalactic objects 
irradiated by ultraviolet background (UVB) radiation in a reionized
universe.
For the purpose, 
we perform three-dimensional radiation smoothed-particle-hydrodynamics (RSPH)
calculations, where the radiative transfer is solved by a direct method and
also the non-equilibrium chemistry of primordial gas including
H$_2$ molecules is incorporated.
Attention is concentrated on radiative transfer effects
against UVB for the formation of subgalactic objects with $T_{\rm vir}\la 10^4$K. 
We consider the reionization model with $z_{\rm reion} \approx 7$
and also the earlier reionization model ($z_{\rm reion} \approx 17$) 
inferred by the WMAP.
We find that the star formation is suppressed appreciably by
UVB, but baryons at high-density peaks are self-shielded even
during the reionization, forming some amount of stars eventually.
In that sense, the photoevaporation for subgalactic systems 
is not so perfect as argued by one-dimensional spherical calculations.
The final stellar fraction depends on the collapse epoch and
the mass of system, but almost regardless of the reionization
epoch. For instance, a few tenths of formed stars are born after the cosmic
 reionization in $z_{\rm reion} \approx 7$ model, while more than 90\% 
stars are born after the reionization in the WMAP reionization model.
Thus, effects of UVB feedback on the substructure problem 
with a cold dark matter (CDM) scenario should be evaluated with careful
 treatment of the radiative transfer.

The star clusters formed at high-density peaks 
coalesce with each other in a dissipationless fashion 
in a dark matter potential, resultantly forming a spheroidal system.
As a result, these low-mass galaxies have large mass-to-light ratios
such as observed in dwarf spheroidals (dSph's) in the Local Group.

\end{abstract}

\section{Introduction}
\label{intro}
In the context of cold dark matter (CDM) cosmology,
the first generation of objects should have the mass of $\sim 10^6M_\odot$
and form at redshifts of $20 \la z \la 50$ \citep{Teg97,FC00}. 
At later epochs, the first objects are assembled into larger systems
in a hierarchical fashion to form dwarf or normal galaxies.
On the other hand, it is thought that the universe was reionized
at $z>5$, because the absorption by the neutral intergalactic medium (IGM)
is observed to be feeble at redshifts $z \la 4$.
The reionization redshift is inferred to be 
$z_{\rm reion} \approx 6$ from
the spectra of high-$z$ quasars (Djorgovski et al. 2001, and
references therein), $6 \la z_{\rm reion} \la 10$ from
simulations on Ly$\alpha$ absorption lines \citep{UNS01},
or $11 \la z_{\rm reion} \la 30$ from
the recent results of WMAP \citep{Kog03}.
Thus, it is anticipated that the UVB directly influences 
the assembly phase of galaxies.
The UVB has several important physical impacts on the galaxy
formation. One of the most important effects is the photoheating.
If a gas cloud is irradiated by UVB, the gas temperature is raised up to
$T\sim 10^4-10^5$K, so that
gaseous systems with virial temperature of $T\la 10^4$ K
can be evaporated owing to the enhanced thermal 
pressure \citep{UI84,DR87,Efs92,BR92,TW96,FT00,Gne00c,SR01,SD03}. 
Such an effect by UVB may reconcile the paradox that 
low mass galaxies are overproduced in CDM cosmology, compared 
with observations \citep{WF91,KWG93,Cole94,Moore99,BGS01}.
The criterion for photoevaporation has been derived from detailed 
hydrodynamic calculations by \citet{UI84} and \citet{TW96}.
However, if the systems are self-shielded against the UVB,
the criterion for photoevaporation is completely changed.
The self-shielding comes from the radiative transfer effects
of ionizing photons.
\citet{TU98} have derived the self-shielding criterion
by solving full radiative transfer for spherical clouds,
to find the critical optical depth of 2.4.
Also, if the system is self-shielded from UVB,
the gas can cool below $10^4$ K by H$_2$ cooling
\citep{ SuKi00}. The complete radiation-hydrodynamic 
calculations have been done for spherical clouds by \citet{Kita01},
where hydrodynamics, radiative transfer, 
and primordial gas chemistry including H$_2$ molecules
are self-consistently incorporated, and thereby the criteria
for cloud collapse have been derived depending on the redshifts.
Very recently, pioneering approaches with three-dimensional 
hydrodynamics have started on this isuue in the context of CDM cosmology.
They are, for example, the simulations by \citet{ RGS02a,RGS02b} with
radiative transfer based on 
the optically thin variable Eddington tensor (OTVET) approximation,
and by Tassis et al. (2003) with optically-thin approximation.

If the nonlinear evolution of density fluctuations
proceed in a hierarchical fashion, two major effects 
by radiative transfer are expected. 
One is the direct effect, that is, the self-shielding of density peaks
that collapse prior to the reionization \citep{Gne00c,NUS01,Raz02,RGS02a,RGS02b}. 
The other is the enhancement of H$_2$ formation
in relatively massive fluctuations which are once
ionized and then self-shielded in the course of
mass accretion. The second effect has been pointed
out by many authors so far \citep{KS92,CGP97,SU00}.
These two effects play a substantial role for
the galaxy formation under UVB, especially if the star formation
process is taken into account in collapsing clouds,
because star formation activities in galaxies appear to be
strongly correlated to cold HI gas with a few $\times
10^3$K \citep{YL97a,YL97b}. 
Therefore, the effects of radiation transfer of ionizing photons
should be carefully treated to elucidate the galaxy formation process
under UVB. 
In this paper, we perform 3D SPH calculations, incorporating radiation transfer
and primordial gas chemistry.
Here, we focus on relatively low mass galaxies, where the dynamics is
likely to be strongly subject to UVB. The evolution of subgalactic systems
is significant also for the formation of massive galaxies,
since the self-shielding against UVB can be responsible to overall SFH 
in the galaxies and therefore resultant galactic morphology
\citep{SU00b}. 

The goal of this paper is to elucidate how 
the photoevaporation of subgalactic systems is caused by UVB
and how the star formation history (SFH) is influenced consequently.
The paper is organized as follows. In \S 2, the numerical methods
are described. In \S 3, the results of simulations are presented.
In \S 4, the SFH is discussed in relation to the inferred
SFH in dwarf galaxies. \S 5 is devoted to the conclusions.

\section{Numerical Methods}
\label{Simulation}
We have performed numerical simulations on the single halo collapse with
SPH particles ($2^{15}$) and dark matter particles ($2^{15}$). 
The halo mass ranges from $M_{\rm halo}\approx 6\times 10^7$ 
to $6\times 10^8 M_\odot$. The initial conditions are set by COSMICS. 
The radiative transfer of the UV radiation field is
solved assuming a single source outside the simulation box, coupled with
the detailed chemistry of primordial gas. We also have taken into
account the numerical ``star formation''. Following subsections are
devoted to describe some details of the scheme.

\subsection{Hydrodynamics and Gravity}
\label{GRAVHYDRO}
Hydrodynamics is calculated by Smoothed Particle Hydrodynamics (SPH)
method. We use the version of SPH by \citet{Ume93} with the modification
according to \citet{SM93}, and also we adopt the particle resizing
formalism by \citet{Thac00}. 
The gravitional force is calculated by a special purpose processor for
gravity calculation, GRAPE \citep{Sugi90}, which can
also accelerate an SPH scheme \citep{Ume-etal93}.
Here, we used the latest version,
GRAPE-6 \citep{Makino02}, which has the speed of 1Tflops.
Eight GRAPE-6 boards are combined with CP-PACS,
which is a massively parallel supercomputer in University of Tsukuba,
composed of 2,048 processing units,
with theoretical peak speed of 614 Gflops  \citep{Iwa98}.
This system is called the Heterogeneous Multi-Computer System (HMCS)
\citep{HMCS}. 
The hydrodynamics, radiative transfer, and chemical reactions
are solved with CP-PACS.
The originally developed software for HMCS can be easily applied 
to any massively parallel computer
connected through TCP-IP with GRAPE-6 boards.
Actually, we also used an alpha-cluster and a PC-cluster instead of
CP-PACS as the host.

The softening length for gravity is taken as $60$pc 
for all SPH and CDM particles. Also,
the minimal size of SPH particle is introduced so as to prevent the
simulation from stopping owing to very short local timescales.
We have tested this scheme by the standard Sod's shock tube and Evrard's
collapse problems, and the results well reproduce the known solutions.

\subsection{Thermal Processes}
\label{CH-fn}
The non-equilibrium chemistry and radiative cooling 
for primordial gas are calculated by the code
developed by \citet{SuKi00}, in which H$_2$ cooling and
reaction rates are taken from \citet{GP98}. The H$_2$ cooling
rate induced by the collision with H atoms is plotted in
Figure \ref{fig:coolfn1} for high ($10^6{\rm cm}^{-3}$) 
and low density ($0.1\,{\rm cm}^{-3}$) cases. 
The cooling functions by \citet{HM79} and
\citet{LS83} are also plotted for reference. They agree with
each other at the high density limit (i.e. LTE), whereas they disagree
significantly with each other at the low density limit. 
We also assessed the contribution of metallic cooling, and 
compared it to H$_2$ and Ly$\alpha$ cooling in Figure \ref{fig:coolfn2}
for the temperature below $10^4$K. 
The cooling rate by metals is evaluated
by the fitting formula in \citet{DM73}, assuming that the relative
abundance between the heavy elements is the same as the solar
neighbourhood. The fractions of electrons and H$_2$ 
are assumed to be the values behind a shock 
with the velocity of $v_{\rm s}> 30{\rm km s^{-1}}$, since the
initially ionized gas traces a similar path as the
gas behind such a shock \citep{KS92,CGP97,SU00}. 
We also plot H$_2$ cooling function assuming the H$_2$ fraction
of $10^{-4}$, which is favored for the clouds with virial
temperature of $10^{3}{\rm K}\la T_{\rm vir}
\la 10^4{\rm K} $\citep{NS99}.\footnote{In this case, the fraction of
electrons can be an order of magnitude smaller than
the high velocity postshock cases. 
Hence, the cooling rate by metals becomes an order of
magnitude smaller, since the cooling is induced by the
collision between electrons and heavy elements.} 
As shown in Fig.{\ref{fig:coolfn2}}, 
for $T \la 8\times 10^3$ K, 
the dominant cooling mechanism is dependent on the metallicity.

In nearby dwarf galaxies, the metal cooling is likely to be important,
because the observed metallicity is $Z/Z_\odot \sim 10^{-2}- 10^{-1}$.
On the other hand, the observations of Ly$\alpha$ absorption systems 
at high redshifts imply that the metallicity of
high-$z$ intergalactic medium (IGM) is at a level of
$Z/Z_\odot \sim 10^{-2}-10^{-3}$ \citep{Son01}. 
Hence, the cooling rate by metals is rather
smaller than that by H$_2$, as long as $10^3{\rm K} \la T \la 10^4
{\rm K}$. 
In the present paper, we focus on the primary star-forming phase in dwarf galaxies
from metal poor gas with the metallicity seen in IGM or more primordial gas.
If one pursues the subsequent recycling phase of interstellar 
medium with SN explosions, it is definitely requisite to include 
the metal cooling. Such a recycling effect is important for the chemical
evolution of galaxies, but the issue is beyond the scope of present paper 
and left in future study. 

It should be also noted that the effects by metals can play a significant
role even at a level of $Z/Z_\odot \sim 10^{-5}$
for the runaway collapse onto a primordial protostar 
\citep{Omukai00, BFCL01, Schneider02},
although here we do not pursue the runaway to a star 
in such dense, cold regions. 
Based upon these arguments, we neglect the cooling by metals 
in this paper, and the thermal processes are calculated
with the chemistry for primordial abundance gas.

\subsection{Radiative Transfer}
\label{RT}
The photoionization rate and the photoheating rate are 
respectively given by 
\begin{eqnarray}
k_{\rm ion}&=& n_{\rm HI}\int_{\nu_L}^\infty\int \frac{I_\nu\left( s \right)}{h\nu} \sigma_{\nu} d\Omega d\nu, \label{eq:rateion} \\
\Gamma_{\rm ion}&=& n_{\rm HI}\int_{\nu_L}^\infty \int \frac{I_\nu\left( s \right)}{h\nu} \sigma_{\nu} (h\nu-h\nu_L)d\Omega d\nu. \label{eq:rateheat}
\end{eqnarray}
Here $n_{\rm HI}$ represents the number density 
of neutral hydrogen, $\sigma_{\nu}$ is
the photoionization cross section, $\nu_{\rm L}$ is the frequency at the Lyman limit, and $\Omega$ is the solid angle. $I_{\rm \nu}(s)$ is
the intensity of the UV radiation with $s$ denoting the distance 
from the source of ionizing photons.
$I_{\rm \nu}(s)$ is determined by solving radiative transfer equation:
\begin{eqnarray}
\frac{d I_\nu}{ds}&=& -n_{\rm HI}\sigma_\nu I_\nu,
\label{eq:radtr}
\end{eqnarray}
where the reemission term is not included, because we employ on-the-spot
approximation. 
The transfer equation (\ref{eq:radtr}) is solved based upon the method
developed by \citet{KB00}, which utilizes the neighbor lists
of SPH particles to assess the optical depth from a certain source to 
an SPH particle. 
Here, we assume a single point source located very far 
from the simulated region, and control the UV intensity 
by specifying the incident flux to the simulation box
as described in section \ref{Setup}.
It is noted that the irradiation from one side 
can overestimate the effects of shadowing, 
and also internal sources, if they form,  might play a significant role 
\citep{RGS02a,RGS02b}.
On the other hand, in Kessel-Deynet \& Burkert's method,
the numerical diffusion of radiation is not negligible. 
This effect tends to reduce the shadowing effect.
These points should be improved in the future.

We parallelize the scheme so that we can implement it on
massively parallel computer systems such as CP-PACS.
For the purpose, particles should be classified into the
subsets which are assigned to PUs. The domain decomposition is
performed as shown in Figure \ref{fig:parallel}. The first step is to
project the particle positions onto the plane 
which is perpendicular to the light
rays of the external single source. Since the position of source 
is assumed to be far away from the simulated region, 
we can approximate the light rays to be parallel. 
The second step is to sort the projected positions 
of particles on the $x-y$ plane. Then, with the results
of coordinate sorting, we generate subsets 
with an equal number of particles, which
are respectively assigned to computational domains, i.e. PUs.
In Appendix, the test calculation using this scheme 
is presented for the propagation of ionization 
front in a uniform media. 
We remark that the present
parallelization scheme is not valid for the multiple sources, because
the domain decomposition is based upon the projection of all particles
onto the plane perpendicular to the source direction. 
Hence, it is necessary to invoke a new technique in order to
include internal sources in the future analysis.

Combining equations (\ref{eq:rateion}),(\ref{eq:rateheat}) and
(\ref{eq:radtr}), we can rewrite the ionization rate and the
photoheating rate as follows:
\begin{eqnarray}
k_{\rm ion}&=& -\frac{d}{ds}\int_{\nu_L}^\infty\int \frac{I_\nu\left( s \right)}{h\nu} d\Omega d\nu \nonumber \\ 
 &=& -\frac{d}{ds}\int_{\nu_L}^\infty \frac{F_\nu\left( s \right)}{h\nu} 
d\nu, \\
\label{eq:rateion2}
\Gamma_{\rm ion}&=&-\frac{d}{ds}\int_{\nu_L}^\infty \int \frac{I_\nu\left( s\right)}{h\nu}(h\nu-h\nu_L)d\Omega d\nu \nonumber \\
 &=&-\frac{d}{ds}\int_{\nu_L}^\infty \frac{F_\nu\left( s\right)}{h\nu}(h\nu-h\nu_L)d\nu.
\label{eq:rateheat2}
\end{eqnarray}
When we need to assess the rates at each grid point
(or each SPH particle), we use the following formula: 
\begin{eqnarray}
k_{\rm ion}\left(i+1/2\right)&=& \frac{1}{s_{i+1}-s_i}\left(\Phi_1(i)-\Phi_1(i+1)\right),\\  
\label{eq:rateion3}
\Gamma_{\rm ion}\left(i+1/2\right)&=&\frac{1}{s_{i+1}-s_i}\left(\Phi_2(i)-\Phi_2(i+1)\right),\\ 
\label{eq:rateheat3}
\end{eqnarray}
where
\begin{eqnarray}
\Phi_1(s)&=&\int_{\nu_L}^\infty \frac{F_\nu\left( s \right)}{h\nu} d\nu,\\
\Phi_2(s)&=&\int_{\nu_L}^\infty \frac{F_\nu\left( s \right)}{h\nu}(h\nu-h\nu_L) d\nu.
\end{eqnarray}
Here $i+1/2$ denotes the grid point at which the rates are assessed,
and $i$ denotes the boundary between $i-1/2$ and $i+1/2$ grids.
In the present simulation, these grids are generated by
the method in \citet{KB00}.
The above formula has an important advantage 
that the propagation of ionization front is
properly traced even for a large grid size with 
optical depth greater than unity ($\tau > 1$)
\citep{KB00,ANM99}. 

\subsection{Star Formation Algorithm}
\label{SFA}
Here the ``star formation'' algorithm used in this simulation is
described. At each time step, we pick up the SPH particles that satisfy
following four conditions: 
(1) ${\rm div}\ \mbox{\boldmath $v$} < 0$, (2)
$\rho/\bar{\rho} > 200$, (3) $T < 5\times
10^3$K, and (4) $y_{\rm H_2} > 5\times 10^{-4}$. 
These conditions look similar to those adopted by \citet{CO93}, 
but the conditions (3) and (4) are different. 
The first condition guarantees
that gas surrounding the particle is infalling.
In addition, if the region is virialized, the density
around the particle should satisfy the second condition.
The conditions (3) and (4) cannot be satisfied
unless the region is self-shielded against UVB and thereby
H$_2$ cooling is effective.
These two conditions are essential for the star formation,
although they are not hitherto taken into account 
in the previous simulations.
Also, the conditions (3) and (4) match
the observations on the HI contents in nearby dwarf galaxies,
which indicate that the presence of cold ($T\sim
10^3$K) HI gas is a good indicator of star formation activities 
\citep{YL97a,YL97b}.

The next step is to define the timescale of ``star formation''.
In order to define the conversion timescale from gas to stars, we use the
following simple and often used formula:
\begin{eqnarray}
 \frac{d\rho}{dt}&=&-\frac{d\rho_{\rm star}}{dt},\\
 \frac{d\rho_{\rm star}}{dt}&=&\frac{c_* \rho}{t_{\rm ff}}, \label{SFR}
\end{eqnarray}
where $t_{\rm ff}$ is the local free-fall timescale, $\rho_{\rm star}$ is
the mass density of star particles, and $c_*$ denotes the parameter of
star formation timescale. Following the above expression, an SPH particle
that satisfies the conditions for star formation 
is converted to a collisionless
particle, after $\Delta t = t_{\rm ff}/c_*$.
Here, we use $c_*=0.1$ for the fiducial model, 
and also investigate $c_*=1$ and $c_*=0.01$ for
several cases.
In the previous numerical simulations on the formation of disk
galaxies, $c_*$ is sometimes set to be $0.05-0.1$ 
with applying the cooling criterion of $10^4$K,
in order to regulate both the star formation efficiency 
and the star formation rate (SFR) to match
the Kennicutt's low (e.g. Kennicutt 1998; Koda, Sofue \& Wada 2000). 
However, in the present simulation, the star formation efficiency
is physically regulated in terms of the temperature criterion (3).
Hence, in this paper, $c_*=1$ does not mean that all the gas cooled to 
$10^4$K is converted into stars. Here, $c_*$ just controls the star 
formation timescale or SFR. 

\subsection{Setup}
\label{Setup}
The initial particle distributions in a CDM-dominated universe 
are generated by a public domain code, GRAFIC,
which is a part of COSMICS. Throughout this paper, we assume
a $\Lambda$-dominated flat universe, 
with $\Omega_{\rm M}=0.3$, $\Omega_{\rm \Lambda}=0.7$, $\Omega_{\rm
B}h^2=0.02$, and $h=0.7$. 
Total mass of the simulated region is $10^8-10^9 M_\odot$, in which
a halo collapses at $1 \la z_{\rm c} \la 10$.
We first generate the positions and velocities
of particles in a cube with the constraint that the peak of
an overdense region is located at the center of the cube. 
The peak height is controlled so that the
overdense region collapses at a given epoch. 
Then, we hollow out a spherical region from the cube, 
so that the sphere contains the
overdense region and the radius of the sphere agrees with the
smoothing scale of the overdensity. 
Then, we attach  a rigidly rotating velocity field to the spherical
region which corresponds to spin parameter $\lambda=0.05$ \citep{HP88,SB95}.
In each run, 
we define the halo mass $M_{\rm halo}(z)$ as the mass within the radius
$R_{\rm vir}(z)$. Virial radius $R_{\rm vir}(z)$ is defined by the condition, \begin{equation}
\bar{\rho}_{DM}(<R_{\rm vir}(z)) = 18\pi^2 \rho_0(1+z)^3,
\end{equation}
where $\bar{\rho}_{DM}(<R)$ is the averaged dark matter density within the radius
$R$ measured from the center of mass of the whole system. $\rho_0$
denotes the averaged density of the present universe.
We also define the collapse epoch $z_{\rm c}$ as the epoch by which a half of
the dark halo mass collapsed within the radius $R_{\rm vir}(z)$.
The typical halo mass just after the virialization is approximately 60\% of the total
mass of the simulated region. These are tabulated in Table \ref{tab:summary}.


For the largest simulations, 
$2^{17}(=131072)$ SPH particles
+ $2^{17}$ dark matter particles are employed. Also, the
case study is done for all parameters  with a smaller number of particles, $2^{15}$ SPH
particles + $2^{15}$ dark matter particles.
Finally, we give the Hubble expansion
velocity to all particles in addition to the already 
generated peculiar velocity fields.

We also fix the initial abundance of chemical species. We have taken the
values of the canonical model in \citet{GP98}. 
The evolution of the ultraviolet background (UVB) is modeled
as follows. The spectrum shape of UVB is assumed to be $I_\nu \propto
\nu^{-1}$, and
the intensity $I_{\rm 21}$, which is the UVB intensity 
normalized by $10^{-21}{\rm erg~ s^{-1}
cm^{-2} str^{-1} Hz^{-1}}$, is assumed to be
$I_{21}=\left[\left(1+z\right)/3 \right]^3$ for $z\le 2$ 
and $I_{21} =1$ for $2 < z \le 4$.
This dependence is consistent with the UV intensity
in the present epoch \citep{Mal93,DS94} and the value inferred from 
the QSO proximity effects at high redshifts 
\citep{BDO88,GCD96},
although there is an uncertainty of $10^{\pm 0.5}$ at $2 < z \le 4$. 
As for higher redshift epochs, we employ two regimes.
One is a regime with $z_{\rm reion} \approx 7$, which is
provided by $I_{21}=\exp\left[3\left(4-z\right)\right]$ for $ z > 4$
\citep{UNS01}.
Such UVB evolution is suggested by the comparison between
Ly$\alpha$ continuum depression in high-$z$ QSO spectra
and the simulations of QSO absorption lines based on
6D radiation transfer calculations on the
reionization \citep{NUS01}. 
In this regime, the reionization proceeds from 
$z \approx 9$ and is completed at $z \approx 7$. 
Most of the runs are performed in this regime.
But, for comparison, 
we also investigate a high-$z$ reionization regime 
($z_{\rm reion} \approx 17$)
such as recently inferred by the WMAP \citep{Kog03}. 
In this regime, the intensity of UVB is modeled by
$I_{21}={\rm max}(\exp\left[3\left(4-z\right)\right], 0.01)$ for $ 4 < z < 17$
and $I_{21}= \exp\left[3\left(17-z\right)\right]$ for $z >17$,
based on \citet{NUS01}. 
We calculate typical two runs in the high-$z$ reionization model.

\section{Results}
\label{Results}

In Table \ref{tab:summary}, the model parameters studied here
and the basic results are summarized, where
$z_{\rm c}$ is the collapse redshift, $M_{\rm halo}$ is the halo mass,
$T_{\rm vir}$ is the virial temperature, and $\sigma_{\rm 1D}$ is
the line-of-sight velocity dispersion of stellar component,
$\sigma_{\rm 1D}({\rm DM})$ is that of dark matter component.
These values should be defined as functions of redshift, but the
tablated values are assessed just after the virialization.
$R_{\rm half}$ is the effective radius of the formed galaxy,
$f_{\rm star}$ is the final stellar fraction, and
$M_{\rm halo}/L$ is the mass-to-luminosity ratio in solar units.
We show the detailed results in the following.

\subsection{ Reionization Feedback }
\label{PEAK}
Figures \ref{fig:2dplot1} and \ref{fig:2dplot2} show two typical
results, where $c_*=0.1$ for both
cases. Figure \ref{fig:2dplot1} shows the
low-$z_{\rm c}$ ($z_{\rm c}\simeq 1.7$) and less massive case ($M_{\rm
halo}\simeq 6\times 10^7 M_\odot$). 
Dots in the figure denote the location of SPH particles and
numerically formed stars. The colors of particles represent the
gas temperature (color legend is shown at the bottom), except that
blue particles denote stars.
At $z\simeq 66$ (1st panel), 
the initial condition is set in a spherical region 
as described in section \ref{Setup}.
Thereafter, the system expands by the cosmic expansion,
and also baryon perturbations are induced by 
imprinted dark matter fluctuations (2nd panel). 
At $z\simeq 9$, first objects with the mass of
$\sim 10^{6-7} M_\odot$ form at density peaks (2nd and 3rd panels).
This is consistent with the previous results 
by \citet{Teg97} and \citet{FC00}.
At $z \approx 7$, the system is reionized overall and reheated by 
UVB (4th panel). 
Then, a portion of the cooled gas that is once settled into the
first objects is evaporated as a pressure-driven thermal wind
at $z \approx 5$ (5th panel). But a significant amount of baryons 
are self-shielded against UVB at high-density peaks 
even during the reionization,
so that the photoheating is precluded deep inside the peaks
and eventually stars form there.
As a result, small star clusters are left after 
the reionization epoch (6th panel).
The stellar clusters coalesce with each other 
in a dissipationless fashion,
according as dark matter fluctuations grow 
hierarchically (7th panel). 
Finally, a subgalactic spheroidal system 
with a high total mass-to-stellar mass ratio 
($M_{\rm halo}/M_{\rm star}=20$) forms.
If we assume the stellar mass to luminosity ratio of 3
in solar units
that comes from a Salpter-like initial mass function,
then the mass-to-light ratio of this system is assessed
to be $M_{\rm halo}/L =63$.

In the above case, baryons are cooled to form stars only 
at density peaks, and gas in other regions is
photoevaporated, resulting in a thermal wind.
But, the evolution is quite different for the high-$z_{\rm c}$ and
massive halo model.
Figure \ref{fig:2dplot2} represents the $z_{\rm c}\simeq 7.6$ and $M_{\rm
halo}\simeq 6\times 10^8 M_\odot$ case. 
In this case, most of baryons collapse before the reionization.
Resultantly, the photoevaporation is less effective
than the low-$z_{\rm c}$ case, and thus the whole system does not 
lose a significant portion of baryonic matter. 
Finally, a formed spheroidal system has $M_{\rm halo}/M_{\rm star}=7.7$,
which is close to the assumed initial the total mass-to-baryonic mass ratio,
$\Omega/\Omega_{\rm B}=7.35$.
The mass-to-light ratio of this system is assessed
to be $M_{\rm halo}/L =26$.

\subsection{Star Formation History}
\label{SFH}
In Figure \ref{fig:sfr}, the star formation histories (SFH)
are shown for typical four runs, which are
the low mass ($6\times 10^7 M_\odot$) case with $z_{\rm c}=1.7$ and 8.1
and the high mass ($6\times 10^8 M_\odot$) case with $z_{\rm c}=1.1$ and 7.6.
As commonly seen, the SFR is peaked before $10^9$yr
($z\ga 5$)
and the star formation activity is substantially suppressed
by UVB after the reionization, $z\la 5$. 
This indicates that the star formation mainly occurs 
in high-density regions which collapsed prior to the reionization.
The peak SFR is lower in the systems with lower collapsing epochs.
However, the difference is not so drastic.
The peak SFR in $z_{\rm c}\approx 1$ cases is roughly a half of that
in $z_{\rm c}\approx 8$ cases. 
If based on the criterion for spherical clouds 
(e.g. Kitayama et al. 2001),
the clouds should be completely photoevaporated in
the two cases of $z_{\rm c}\approx 1$.
But, the present results show that high-density peaks
which collapse before the reionization contribute significantly to 
the star formation even in such low-$z_{\rm c}$ objects.
This is a consequence of the hierarchical growth of
density fluctuations that is caused by
the CDM spectrum.

As another important result, it should be noted that 
the star formation activity in
the case with $M_{\rm halo}\simeq 6\times 10^8 M_\odot$ for high-$z_{\rm c}$
continues even after the reionization.
This is caused by the depth of gravitational potential. 
A massive galaxy with high-$z_{\rm c}$
has a deeper gravitational potential. Thus, 
the dark halo can prevent the photoheated
gas from blowing out into the intergalactic space. 
Instead, the heated gas slowly accretes in the potential of dark halo.
Eventually, the gas is self-shielded from UVB and cools by
H$_2$, so that stars form even in UVB.

In Figure \ref{fig:f_star}, the final fraction of stars,
$f_{\rm star}$, is plotted against collapse epoch $z_{\rm c}$,
where $f_{\rm star}$ is defined by
the total mass of star particles divided by
the total baryonic mass at $z=1$
($f_{\rm star}\equiv M_{\rm star}/M_{\rm baryon}$).
The starred pentagons represent the results of $M_{\rm halo}\simeq 6\times 10^{8}M_\odot$
runs, and the pentagons denote those of $M_{\rm halo}\simeq 6\times 10^{7}M_\odot$ runs, where $c_*$ is assumed to be $0.1$.
The stellar fraction is smaller for $M_{\rm halo}\simeq 6\times 10^7M_\odot$ cases than
$M_{\rm halo}\simeq 6\times 10^8 M_\odot$ cases. This comes from the SFH shown above.
The massive objects tend to retain more gas components 
than less massive objects in the presence of UVB.

Additionally, $f_{\rm star}$ decreases monotonously
with decreasing $z_{\rm c}$ for
both $M_{\rm halo}\simeq 6\times 10^8 M_\odot$ and $M_{\rm halo}\simeq 6\times 10^7 M_\odot$ cases. 
It is found that the dependence of $f_{\rm star}$ on $1+z_{\rm c}$ 
is almost linear, and its relation is given by
\begin{equation}
f_{\rm star}\simeq a(M)\left(1+z_{\rm c}\right)+b(M)
\label{fstar}
\end{equation}
where 
$a(6\times 10^8 M_\odot)=7.7\times 10^{-2}$, $a(6\times 10^7
M_\odot)=6.9\times 10^{-2}$ , 
$b(6\times 10^8 M_\odot)=0.26$ and $b(6\times 10^7 M_\odot)=7.7\times10^{-2}$.
We emphasize that even for galaxies assembled after the reionization, 
some fraction of baryons are cooled and form stars at high-density peaks. 
The vertical two lines denote the critical redshifts obtained by \citet{Kita01}, 
after which protogalactic clouds are photoevaporated. 
The condition for $M_{\rm halo}\simeq 6\times 10^7 M_\odot$
is denoted by a solid line
and that for $M_{\rm halo}\simeq 6\times 10^8 M_\odot$
by a dotted line.
This result shows again that the inhomogeneity in protogalactic clouds
is crucial for the star formation in them,
because a portion of baryons collapse and cool before the reionization,
and also some fraction of gas
is self-shielded from UVB even after the reionization
if the gas accretes in dark matter potential.

In Figure \ref{fig:f_star}, we plotted also the results with
$c_*=1$ (high SFR runs) and $c_*=0.01$ (low SFR runs). 
As seen clearly, the results are basically the same
as the case of $c_*=0.1$.
This is physically understood as follows: 
In the present simulations, stars form from overdense regions 
with $\rho\ga 10^{-22} {\rm g~cm^{-3}}$. 
Then, the star formation timescale 
assessed by the present star formation algorithm (\ref{SFR})
is given by $t_{\rm SF}= t_{\rm ff}/ c_*
\la 10^7 ~c_*^{-1} {\rm yr} $. 
On the other hand, the UV feedback is quite effective 
at $z\la 5$ which corresponds to
$t_{\rm H}\ga 10^9 {\rm yr}$. Thus, the main episode of 
the initial star formation ends before the reionization 
as far as $c*\ga 0.01$. 
Moreover, the density of star forming regions is
higher if $c_*$ is smaller, 
because further radiative cooling works before the gas
is converted to stars. Therefore, the local
free-fall time of star forming region is shorter for smaller $c_*$.
Consequently, $c_*$ is not so important to determine 
the final stellar fraction. 
However, if we adopt $c_*=0.01$, the star formation history is 
modified as a matter of course. The epoch of star formation
is shifted to later epoch for $c_*=0.01$, as shown in 
Fig.\ref{fig:vlowsf_sfr}. Especially, in the cases with high mass 
and high-$z_{\rm c}$, the star formation activity after 
the reionization is enhanced compared to the $c_*=0.1$ case. 
The relation of this result to the continuous star formation history 
inferred in nearby dwarf galaxies (e.g. Mateo 1998) will be discussed
later.

\subsection{Radiative Transfer Effect}
Here, we show more specifically the effects of radiative transfer
in the present simulations. 
For comparison, we perform the control runs in which gas is 
assumed to be optically thin against UV radiation. 
In Figure \ref{fig:f_star_evol}, the time variations of stellar fractions
are compared between the runs with and without radiative transfer.
The parameters are the same as those in Figure \ref{fig:sfr}.
For all runs, it is clear that 
the star formation activities are supressed at earlier epochs
and thoroughly terminated after the reionization in optically-thin simulations.
In Figure \ref{fig:f_star_thin},
the final stellar fractions ($f_{\rm star}$) are shown for 
the runs with and without radiative transfer. 
The optically-thin approximation underestimates 
the stellar fraction by a factor of 1.5-2 for all runs in
the $z_{\rm reion}\simeq 7$ models,
regardless of the mass and collapse epoch.
The radiative transfer effect is even more crucial
in $z_{\rm reion}\simeq 17$ models as shown in \S \ref{WMAP} below.
The reduction of $f_{\rm star}$ by dismissing the radiative transfer
is more than an order of magnitude if $z_{\rm reion}\simeq 17$.
Hence, it is clear that the self-shielding does work to form
stars effectively during the reionization.


Also, in the bottom-right panel 
of Figure \ref{fig:f_star_evol}, 
where $M_{\rm halo}\simeq 6\times 10^8 M_\odot$ and $z_{\rm c}=7.6$,
$f_{\rm star}$ continues to increase if the radiative transfer is incorporated.
Since the gravitational potential is as deep as to
sustain the photoheated gas in this case, some gas is self-shielded in the
course of accretion onto local density peaks, so that stars continue to form.
In the optically-thin simulation, the accreted gas cannot
be cooled below $10^4$K owing to the absence of shielding effect,
so that no stars form after the reionization.

\subsection{Kinematic Properties}
\label{Kinematics}
In Table \ref{tab:summary}, 1D velocity dispersion of the formed
stars ($\sigma_{\rm 1D}$) and effective size $R_{\rm half}$ of the
resultant spheroidal system are listed for various runs. 
Here, the effective size, $R_{\rm half}$, is defined 
so that a half of stars are contained within the radius.
In the low-$z_{\rm c}$ case shown in Figure \ref{fig:2dplot1}, gas can
cool at density peaks before the reionization 
and form compact star clusters. The star clusters coalesce
in a dissipationless fashion to form a single spheroidal system eventually. 
On the other hand, in the high-$z_{\rm c}$ case as shown in Figure
\ref{fig:2dplot2}, the formation process of the final stellar system is
rather different, because the star formation proceeds not
only at the density peaks but also after the collapse of the whole system,
which takes place after the reionization. 
In order to clarify the difference quantitatively, we plot the
time variation of the ratio of the total kinetic energy by random motion 
($E_{\rm ran}$) to the total rotation energy ($E_{\rm rot}$). 
Here, $E_{\rm rot}$ is defined as the summation of 
the rotational energy of
baryonic particles (i.e. SPH particles and star particles) which is
contained in a sphere with radius of $R_{vir}(z)$ from
the center of gravity;
\begin{equation}
E_{\rm rot}=\sum_{r_{\rm i} < R_{\rm vir}(z)} \frac{1}{2}m_{\rm i}
v_{\rm rot,i}^2,
\end{equation}
where $v_{\rm rot,i}$ is the angular velocity of {\it i}-th particle.
$E_{\rm ran}$ is defined by the rest of the kinetic energy, that is,
$E_{\rm ran}=E_{\rm kin}-E_{\rm rot}$, where $E_{\rm kin}$ is 
the total kinetic energy.
When the velocity field is isotropic, the ratio $E_{\rm rot}/E_{\rm
ran}$ should be $1/5$ according to the present definition.
Figure \ref{fig:ratio} shows the time evolution of the ratio $E_{\rm
rot}/E_{\rm ran}$ for typical four runs after the collapse epoch.
As shown in this figure, $E_{\rm rot}/E_{\rm ran}$ is well below 
unity for all the cases after $z_{\rm c}$.  This means
that these systems are not rotationally supported. 
However, if we take a closer look, the ratio for the
high-mass and high-$z_{\rm c}$ case 
($M_{\rm halo}\simeq 6\times 10^8 M_\odot, z_{\rm c}\simeq 7.6$)
is somewhat larger than the others. This reflects
the fact that a part of gas forms a rotating disk before it
is converted into stars.
On the other hand, for the other cases, gas is lost
after the reionization era ($z \la 7$), 
and no rotating disc forms. 

\subsection{Effects of Early Reionization}
\label{WMAP}
Here, we attempt to evaluate the effects of early reionization 
as inferred by the results of WMAP \citep{Kog03}. We perform two runs in
the high-$z$ reionization regime ($z_{\rm reion} \approx 17$)
described at the end of \S \ref{Setup}. The both runs are 
high-$z_{\rm c}$ models, where $M_{\rm halo}\simeq 6\times 10^7 M_\odot$ 
with $z_{\rm c}=8.1$ 
or $M_{\rm halo}\simeq 6\times 10^8 M_\odot$ with $z_{\rm c}=7.6$.
The star
formation history and the time evolution of stellar fractions are shown
in Figure \ref{fig:highz_sfr}. First, for the former run 
($M_{\rm halo}\simeq 6\times 10^7 M_\odot$),
the effects of early reionization feedback are noticeable, and the fraction of
stellar component becomes roughly four times smaller than the
fiducial reionization regime ($z_{\rm reion} \approx 7$). 
Nonetheless, some fraction of baryons
are cooled to form stars even at $z < 17$, owing to the effects of
self-shielding. This is partly because the intergalactic medium is not
perfectly transparent to UVB, even if the UV intensity is 
strong enough to reionize the universe \citep{NUS01}.
On the other hand, for the latter run
($M_{\rm halo}\simeq 6\times 10^8 M_\odot$), the final
stellar fraction is smaller than the fiducial model, but the effect is
not so drastic as the former case. In this case, the final fraction of
stars is $\sim 20 \%$ smaller than the fiducial model.

We also stress that the effect of radiation transfer is very important
for such early reionization model. In the lower panels of Figure
\ref{fig:highz_sfr}, the results from the optically-thin simulations are
also plotted by dashed curves. It is quite clear that the star formation 
is almost prohibited if the optically thin is assumed,
where the final stellar fraction is reduced by an order of magnitude.
The results of more systematic investigation will be reported 
in a forthcoming paper.

\section{Discussion}

\subsection{Substructure Problem}

Recent numerical simulations predict that
subgalactic dark halos are overabundant compared to
dwarf spheroidals observed around our Galaxy 
\citep{Moore99}.
So far, the possibility that this substructure problem 
might be reconciled if the star formation is significantly
suppressed by UVB in subgalactic halos
has been discussed extensively.
However, the present simulations have shown that 
the suppression by UVB is not complete.
Especially, even in the early reionization regime, 
the photoevaporation by UVB 
is not complete and the effects of self-shielding
are still significant.
Hence, in the context of CDM cosmology,
some fraction of baryons forms stars inevitably.
Thus, we have to take into account the shelf-shielding properly, in order
to evaluate the effects of UVB feedback on substructure problem.

To make a further approach on this problem, other effects such as 
the internal UV sources, the evaporation driven by SN explosions 
with the strongly top-heavy initial mass function of formed stars
or the gas tripping in intracluster medium may have to be
investigated.

\subsection{ dE/dSph's in Local Group }

What objects correspond to the simulated galaxies? 
One possibility is dE/dSph galaxies in the Local
Group (LG). The dwarf galaxies in LG are known to be extended (low surface
brightness), spheroidal system (e.g. Mateo 1998). Especially, faint
dSph's ($M_V \ga -14$) have a larger mass-to-light ratio.
\citet{Hirashita98} have found 
\begin{equation}
\log (M_{\rm vir}/L)=2.0 \log M_{\rm vir} + 13.1
\end{equation}
for $ M_{\rm vir}<10^8 M_\odot$, while
$\log (M_{\rm vir}/L)\simeq 0.4-0.7$ for $ M_{\rm vir}>10^8 M_\odot$.
These observational features indicate 
that the mass loss is more prominent for $ M_{\rm vir}<10^8 M_\odot$.
This can be attributed to SN-driven winds 
\citep{Hirashita98,MFM02}.
The present numerical simulations suggest that such
a wind can occur also by the photoheating due to the reionization, 
even before SN explosions drive a thermal wind.
If we use the final stellar fraction (\ref{fstar}),
we find $M_{\rm halo}/L=63$ with assuming $M_{\rm star}/L=3$
for $z_{\rm c}=1.7$ and $M_{\rm halo}\simeq 6\times 10^7 M_\odot$. Such a high M/L
is also obtained by the recent results by \cite{Ricotti03} in which 3D
cosmolgical radiation hydrodynamical simulations are performed with
OTVET approximation \citep{GA01}. We also remark that \cite{SD03}
pointed out the connection between
dSphs and photoevapolated galaxies with 1D simulations.

In addition, the SFH in the Local Group dwarfs has a ubiquitous feature
(e.g. Mateo 1998; Gnedin 2000a). The stars in most LG dwarfs are estimated to
have formed more than 10 Gyrs ago. This means that star formation rate should
have a clear cut-off at such epoch. If we assume the age of the universe
is 13 Gyr, 10 Gyr in lookback time corresponds to $z\sim 3$, 
which is 2 Gyrs later than the cut-off epoch ($z\sim 4-5$) in the present simulations.
However, the observed star formation history itself should
have uncertainties of a few Gyrs.
Thus the reheating by reionization could be one
of the promising mechanism to provide such a cut-off in SFH 
of the LG dwarf galaxies. 
Also, the SFH in the LG dwarfs has another important
feature: there still remain weak star formation activities after the
cut-off of star formation rate. This feature may be reproduced by
the numerical galaxies presented here, 
if $c_*$ is at a level of 0.01 for relatively high mass 
and high-$z_{\rm c}$ models (see \S 3.2). 
Furthermore, if one incorporates the recycling of gas ejected from the stars, 
which is not included in the present simulation, additional stars may form
unless the compete evaporation does not occur.
This tendency is expected to be conspicuous for the halos 
with a deep gravitational potential.
However, the gas ejected to the intersteller space in such dwarf galaxeis
will be photoionized again. The effects
of self-shielding and the formation of dust will delay the
photoevaporation of the processed gas 
and consequently some amount of gas will be converted into stars again. 

\subsection{ dIrr's in Local Group}

On the other hand, the observed dIrr's in LG still have significant star
formation activities embedded in old extended spheroidal components.
Although we have no counterpart in the present simulations, we can infer the
origin of such galaxies by extrapolating the numerical results.
Massive ($\sim 10^{10} M_\odot$) halos formed at low $z_{\rm
c}$ might be the candidate of such galaxies. In
such halos, small scale perturbations grows slowly.
Thus, the fraction of gas component
 after the reionization of the universe is large as seen in Figure
 \ref{fig:f_star}. In addition, such halos have
a gravitational potential deep enough to sustain photoionized gas,
 so that stars form efficiently from such abundant gas. 
Therefore, active star formation is expected
 even after the reionization.
We can suggest that the conditions requisite for dIrr formation
are 1) that the gravitational
potentials of dark halos are deep enough to retain photoheated gas, and 
2) that the galaxy formation epoch
is later than the reionization epoch.
Actually, dIrr's appear to be rather massive (e.g. Mateo 1998).
Also, the reduction of star formation rate due to UVB may suppresses
the SN-driven galactic wind and thereby allows the formation of
diffuse dwarf systems like Irr's.

\section{Conclusions}

In this paper, 
we have performed 3D radiation hydrodynamic simulations on the
formation of low-mass objects under UVB. 
The suppression by UVB of the formation of low-mass objects at lower redshifts
($M \la 10^9 M_\odot, z_{\rm c} \la 5 $) is confirmed.
But, simultaneously it is found that the 
formation of low-mass galaxies at low redshifts is not completely
 forbidden by the UVB-induced photoevaporation. 
Baryons at high-density peaks can collapse and
be self-shielded from UVB even during the reionization,
to form stars eventually.
This occurs even if the universe was reionized at earlier epochs
as suggested by the WMAP.
Thus, effects of UVB feedback on the substructure problem 
with a cold dark matter (CDM) scenario should be evaluated with careful
 treatment of the radiative transfer.

In order to clarify the effects of radiative transfer,
we have compared the results of simulations with those under
the assumption of optically-thin medium.

The reduction for the final stellar fraction is 30-40\% without
radiative transfer in the $z_{\rm reion}\simeq 7$ regime. 
Hence, the star formation before the reionization is important
to determine the final stellar fraction. 
But, the radiative transfer effect is much more serious 
in the earlier reionization ($z_{\rm reion}\simeq 17$) regime, 
where the reduction for the final stellar fraction 
is more than an order of magnitude.
Also, it should be stressed that 
in the optically-thin approximation, gas
temperature of the cloud becomes always higher than $10^4$ K after the
reionization, so that stars cannot form there, whereas
the self-shielding arising from radiative transfer
allows star formation even after the reionization
if the gravitational potential is deep enough to retain the gas.
The star formation after the reionization must influence 
the subsequent evolution of dwarf galaxies 
with the possible recycling of interstellar matter.

In the previous optically-thin simulations 
(e.g. Tassis et al. 2003), the star formation
criterions are different from those in the present simulation. 
The conditions used here are stricter than those used in the previous
works. In previous simulations,
 stars can form slowly even after the
reionization due to the milder star formation conditions, although the
cold HI regions cannot be formed. 
Although such a treatment of star formation might
be approximately valid, it should be confirmed by the
simulations with radiative transfer.

We also have investigated the effects of $c_*$, and found that the
final stellar fractions are not so different for various values of $c_*$.
However, a smaller value of $c_*$ delays star formation and allows
enhanced star formation after the reionization if the system is
relatively massive and the collapse epoch is considerably later
the reionization.
Also, if the stellar feedback is taken into account, 
the further effects of $c_*$ may be expected,
since 1) larger $c_*$ results in faster recycling
(i.e. a larger amount of heavy elements), and 2) larger $c_*$ is followed by
simultaneous SN explosion, which means easier disruption of host galaxies.
These issue should be investigated by the simulations that take into
account the stellar feedback coupled with radiative transfer.

The star clusters formed at high-density peaks coalesce
with each other in a dissipationless fashion to form a spheroidal system
with a large mass-to-light ratio.
It is also found that there is a sharp cut-off of star formation rate at
$z \simeq 4-5$ in all the simulations because of the appreciable
photoevaporation of cold gas clouds in the reionized universe.
Moreover, in massive dwarf galaxies ($M_{\rm halo}\simeq 6\times 10^8 M_\odot$) 
formed at high redshifts ($ z_{\rm c} \sim 8 $), 
weak star formation activities continue even after the
reionization. On the other hand, less massive 
($M_{\rm halo}\simeq 6\times 10^7 M_\odot$) galaxies formed later ($ z_{\rm c}\la 8$) 
do not have star
formation activities after $z \simeq 4-5$. 
Observational counter parts of these systems might be the dwarf spheroidal
galaxies in the Local Group. Those galaxies have extended,
old and spheroidal stellar systems, with characteristic star formation
histories. It is also known that fainter galaxies have larger
mass-to-light ratio. All of those features agree qualitatively with
those of the simulated galaxies, although the stellar feedback
should be also investigated to make quantitative comparison
with observed galaxies.

\acknowledgments

We thank the anonymous referee for helpful comments.
We also thank A. Ferrara, R. Schneider, J. Silk, K. Wada and R. Nishi for
stimulating discussion. 
The HMCS has been developed in a project which Center for Computational
Physics, University of Tsukuba propelled in the course of JSPS
Research-for-the-Future program of Computational Science and
Engineering.
We thank T. Boku, J. Makino and A. Ukawa for cooperative supports
for using the HMCS. 
The analysis has been made with computational facilities 
at Center for Computational Physics in University of Tsukuba. 
We acknowledge Research Grant from Japan Society for the Promotion of
Science (13740124).


\begin{center}
{\small APPENDIX \\
TESTS FOR NUMERICAL SCHEME}
\end{center}
Here, we show some tests of the present numerical scheme.
Using the above scheme, we tested the propagation of
ionization front in a uniform media. The test results are shown in
Figs.\ref{fig:ion} and \ref{fig:temperature}, in which several snapshot of the ionization
structures and temperature distributions are shown. In this test, uniform cube is irradiated from the
left by the ionization flux which is proportional to $\nu^{-1}$. 
The initial (neutral) optical depth of each SPH particle at the Lyman
limit is $\sim 10^3$. 
The vertical axis denote the fraction of electrons (Fig.\ref{fig:ion})
/ temperature (Fig.\ref{fig:temperature}),
and the horizontal axis shows the normalized size of the cube. The
position of the particles are located at the grid points. The axes of
the grid are set so that they are not parallel or perpendicular to the
surfaces of the cube. The vertices denote the ionization structure at
four different times by the RSPH code, and the four solid lines denote
the results obtained by 1D mesh code, which uses $2\times 10^5$ meshes (this
number of mesh corresponds to $\delta \tau \sim 0.1$ at Lyman limit, that is enough to resolve the ionization front). It is
quite clear that RSPH scheme can reproduce the propagation of ionization
structure fairly well, although the optical depth of an SPH particle is
much larger than unity.



\begin{deluxetable}{cccccccc}
\tablecaption{Properties of Formed Galaxies\label{tab:summary} }
\tablehead{
\colhead{$z_{\rm c}$}& \colhead{$M_{\rm halo}/M_\odot$}&
\colhead{$T_{\rm vir}/$K} & \colhead{$\sigma_{\rm 1D}/{\rm km~s^{-1}}$} & \colhead{$\sigma_{\rm 1D}({\rm DM})/{\rm km~s^{-1}}$} &\colhead{$R_{\rm half}/{\rm pc}$} & \colhead{$f_{\rm star}$} & \colhead{$M_{\rm halo}/L$}}
\startdata
1.7&$5.9\times 10^7$&$3.6\times 10^3$&$4.6$&$5.5$&$62$ & 0.28 & 63\\
3.4&$5.9\times 10^7$&$5.1\times 10^3$&$5.1$&$6.5$&$58$ & 0.37 & 50\\
5.1&$6.1\times 10^7$&$6.3\times 10^3$&$5.7$&$7.2$&$69$ & 0.47 & 44\\
6.5&$6.1\times 10^7$&$8.2\times 10^3$&$6.4$&$8.2$&$66$ & 0.61 & 36\\
8.1&$6.1\times 10^7$&$1.0\times 10^4$&$7.1$&$9.2$&$62$ & 0.71 & 33\\
1.1&$6.0\times 10^8$&$1.7\times 10^4$&$14$&$12$&$2.9\times 10^2$& 0.43 & 40\\
2.9&$6.0\times 10^8$&$2.0\times 10^4$&$16$&$14$&$2.1\times 10^2$& 0.55 & 31\\
4.5&$6.0\times 10^8$&$2.0\times 10^4$&$17$&$16$&$2.2\times 10^2$ & 0.68 & 31\\
6.0&$6.1\times 10^8$&$2.1\times 10^4$&$19$&$19$&$1.8\times 10^2$ & 0.79 & 29\\
7.6&$6.1\times 10^8$&$2.6\times 10^4$&$20$&$21$&$1.6\times10^2$ & 0.93 & 26
\enddata
\end{deluxetable}


\setcounter{figure}{0}

\begin{figure}
\begin{center}
\includegraphics[height=10cm]{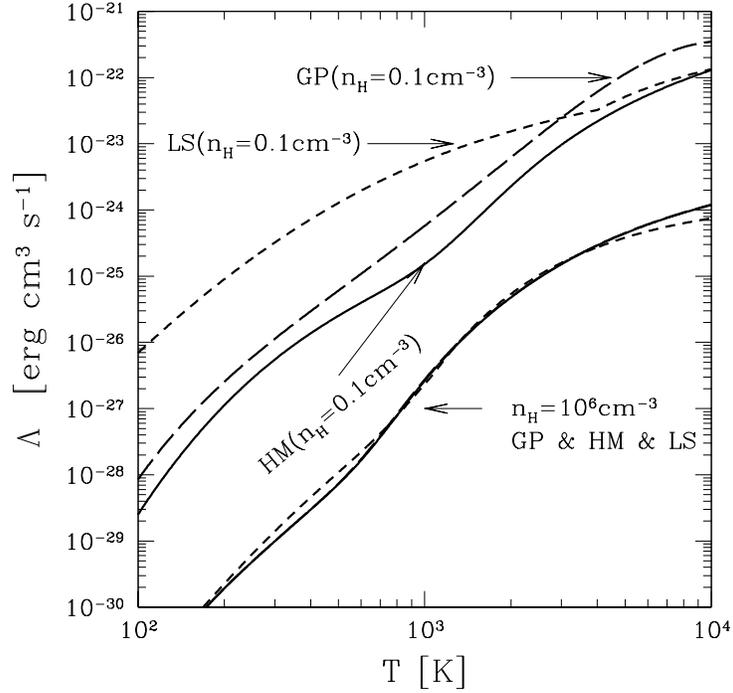}
\caption[dummy]{H$_2$ Cooling functions induced by H-H$_2$ collision are
 plotted against the temperature for two different densities. 
Solid curves are plotted based upon the formula given by
 \citet{HM79}, long dashed curves by \citet{GP98}
 and the short dashed curves by \citet{LS83}.
They are discrepant at low density. In this paper,
those by \citet{GP98} are adopted.}
\label{fig:coolfn1}
\end{center}
\end{figure}

\begin{figure}
\begin{center}
\includegraphics[height=10cm]{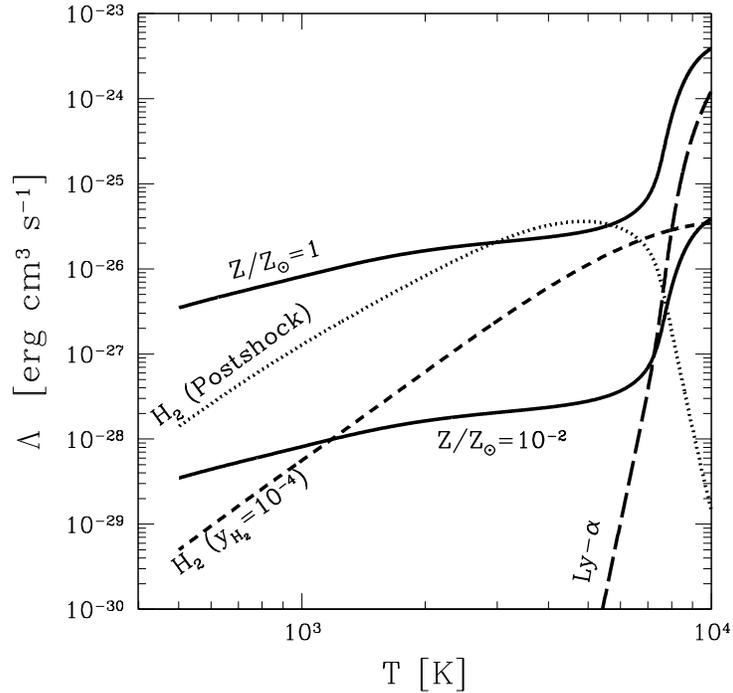}
\caption[dummy]{Cooling functions below $10^4$K are plotted. Two solid lines
 denote the cooling rate due to metals, assuming $Z/Z_\odot=1$ and
 $10^{-2}$ respectively. H$_2$ cooling function is plotted 
for the postshock region with shock velocity larger than 30 km/s 
({\it dotted line}) 
 and also for a given abundance of $y_{\rm H_2}=10^{-4}$ 
({\it short dashed line}). A long dashed line
 represents the rate by hydrogen Ly $\alpha$.}
\label{fig:coolfn2}
\end{center}
\end{figure}

\begin{figure}
\begin{center}
\includegraphics[width=7cm,angle=-90]{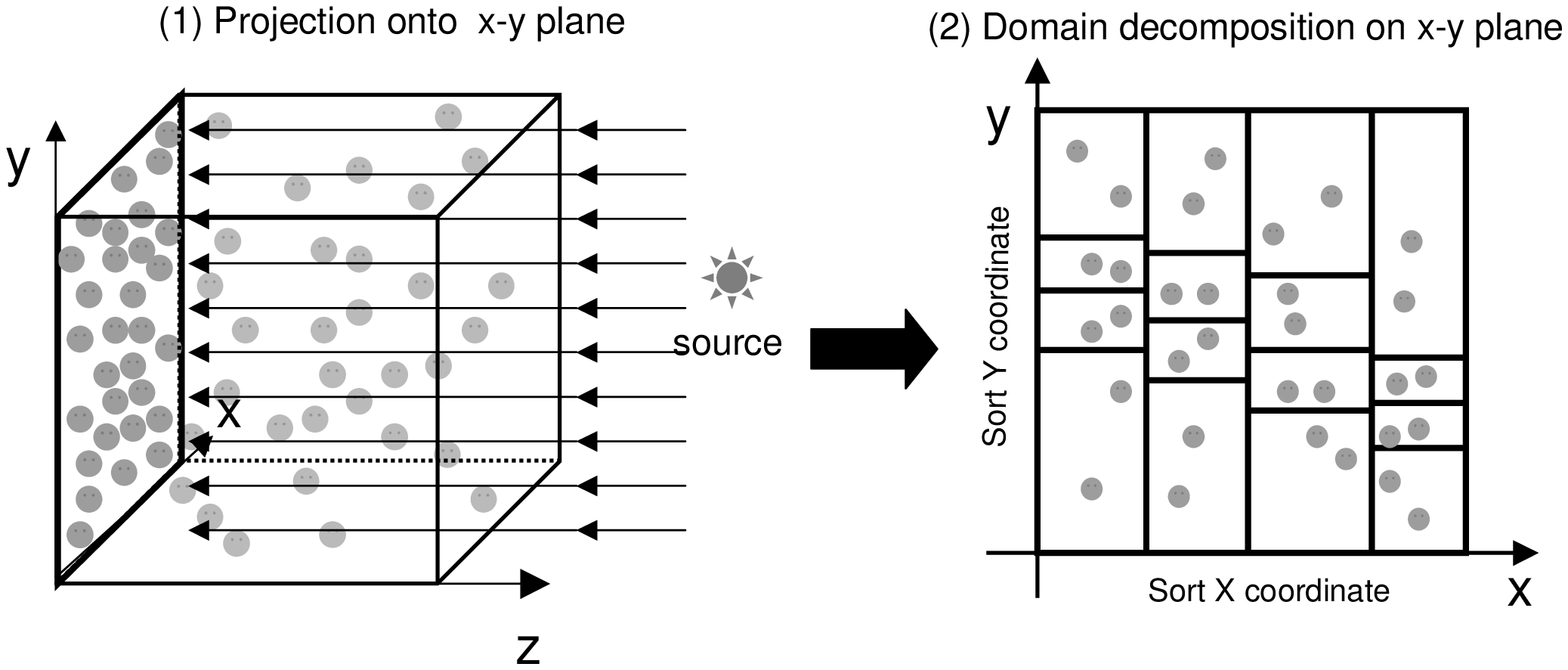}
\caption[dummy]{Parallelization of the code is schematically shown. (1)
 Particles are projected onto the $x-y$ plane which is perpendicular to
 the light ray. (2) The particles are sorted by $x-$ and $y-$
 coordinates and decomposed into subclasses
which include an equal number of particles. 
(3) Each subclass is assigned to each PU.}
\label{fig:parallel}
\end{center}
\end{figure}

\begin{figure}
\begin{center}
\includegraphics[height=20cm]{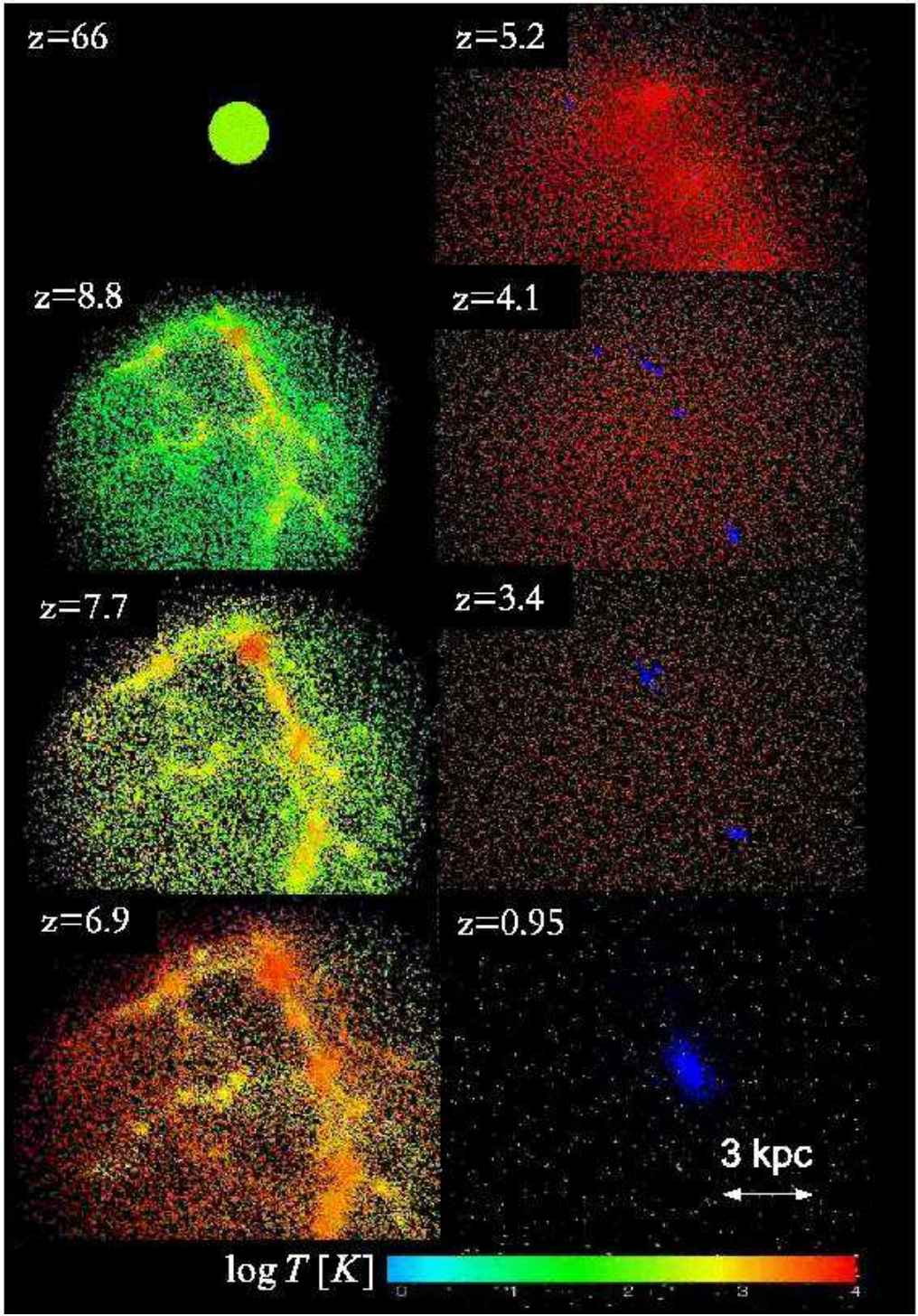}
\caption[dummy]{Time evolution of the spatial distributions of SPH and star particles
 for $M_{\rm halo}\simeq 6\times10^7 M_\odot$ and $z_{\rm c}\simeq 1.7$
 is shown. The figure consists of 8 panels. Upper-left panel is the
 distribution of SPH particles at very early phase, and it evolves downward in the left
column, continuing to the upper-right panel. 
Corresponding redshift is printed at the upper-left corner of each panel.
The colors of dots represent the logarithmic temperature of the SPH particles,
and the color legend is shown at the bottom in logarithmic scale, as well as the physical
 scale ruler (3 kpc) at the right-bottom corner. The blue particles
 do not represent SPH particles, but star particles.}
\label{fig:2dplot1}
\end{center}
\end{figure}

\begin{figure}
\begin{center}
\includegraphics[height=20cm]{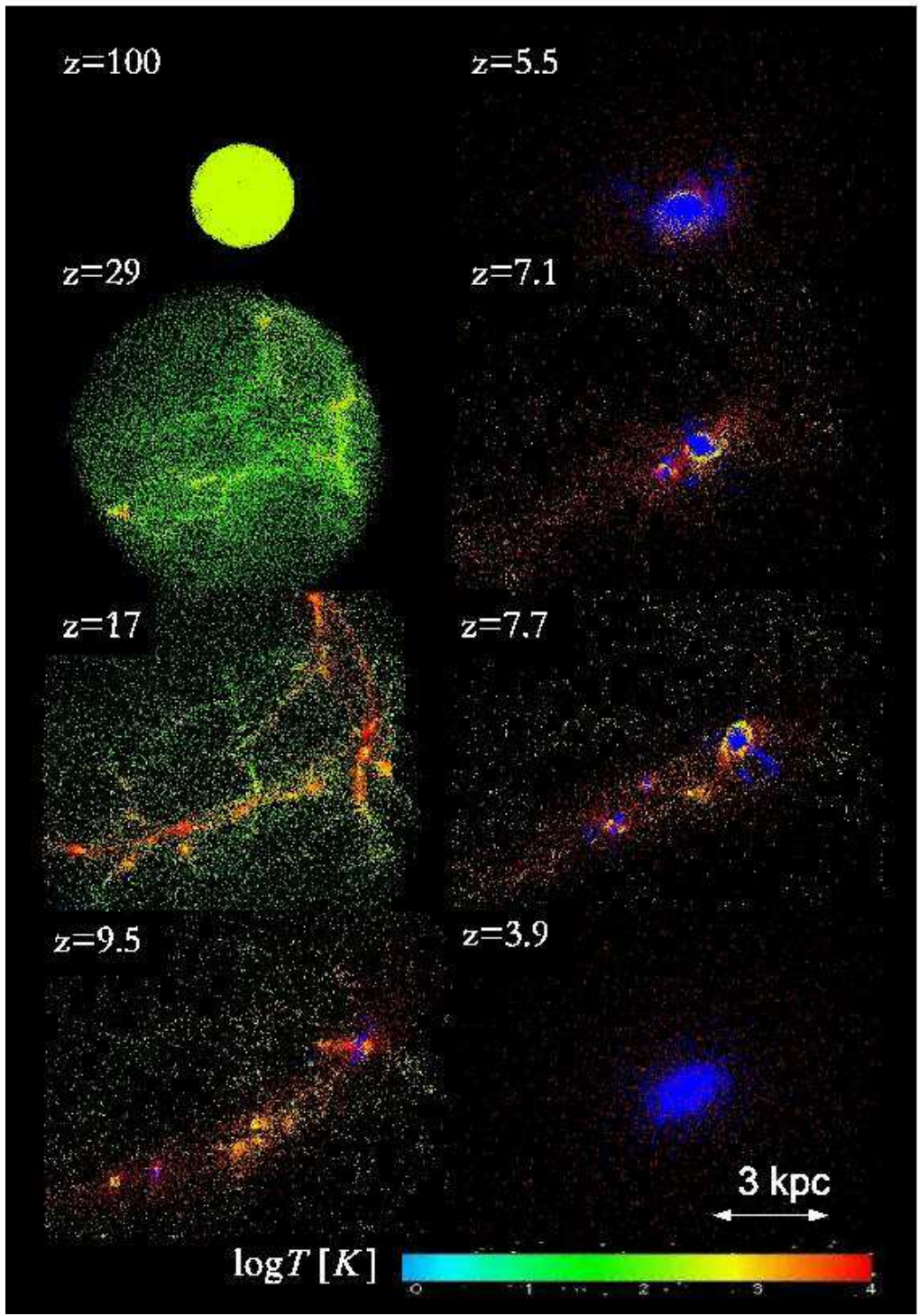}
\caption[dummy]{Same as Fig.\ref{fig:2dplot1} except that the parameters
 are $M_{\rm halo}\simeq 6\times 10^8 M_\odot$  and $z_{\rm c}\simeq 7.6$.}
\label{fig:2dplot2}
\end{center}
\end{figure}

\begin{figure}
\begin{center}
\includegraphics[height=10cm]{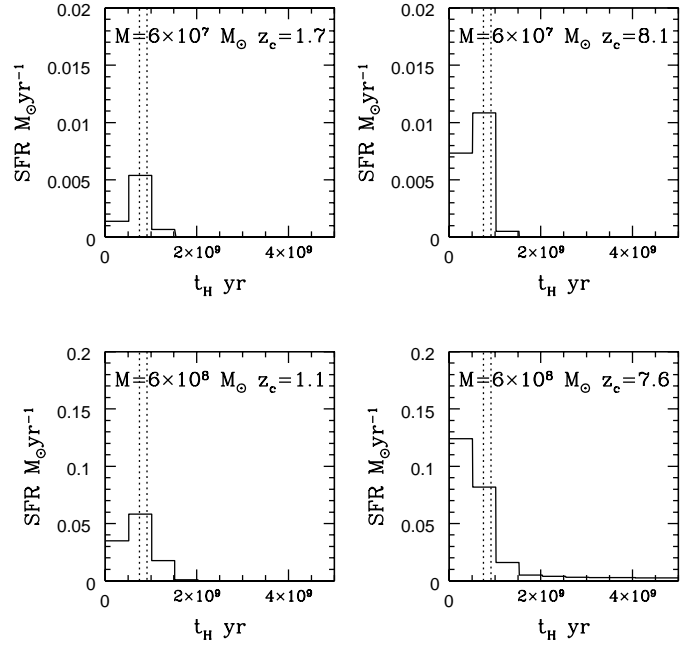}
\caption[dummy]{ The time-variation of star formation rate (SFR)
is plotted against the cosmic time. Four panels represent respectively
the results of four different runs, 
where the parameters are displayed in each panel. The vertical short
 dashed lines denote $z=7$ and $z=6$, between which the reionization is almost
 completed in this model.
 }
\label{fig:sfr}
\end{center}
\end{figure}

\begin{figure}
\begin{center}
\includegraphics[height=10cm]{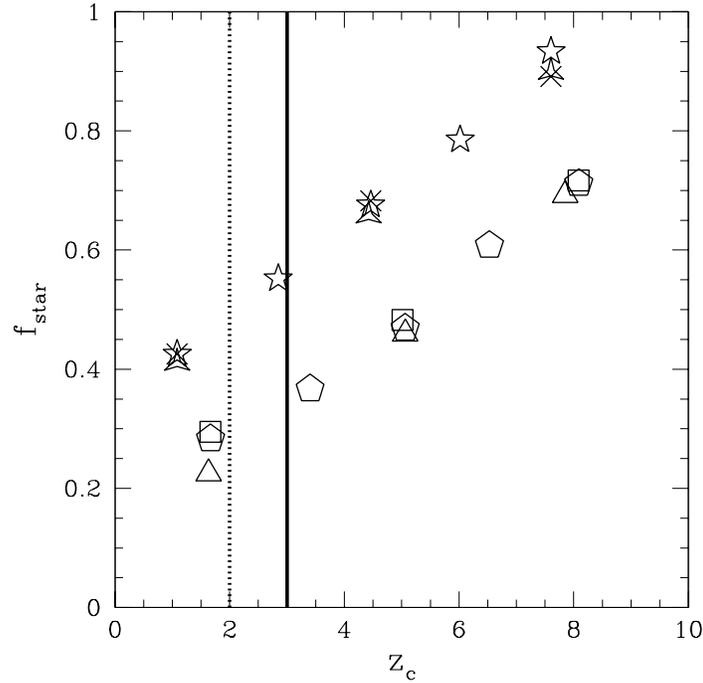}
\caption[dummy]{The final stellar fraction $f_{\rm star}$ is plotted against 
the collapse redshift. Each mark denotes the result of a run: pentagons($c_*=1$),
 squares($c_*=0.1$), and triangles($c_*=0.01$) denote the runs with
 $M_{\rm halo}\simeq 6\times 10^{7}M_\odot$. Similarly, starred pentagons($c_*=1$),
 starred squares($c_*=0.1$), and starred triangles($c_*=0.01$) represent the
 results of relatively massive galaxies ($M_{\rm halo}\simeq 6\times 10^{8}M_\odot$).
}
\label{fig:f_star}
\end{center}
\end{figure}

\begin{figure}
\begin{center}
\includegraphics[height=10cm]{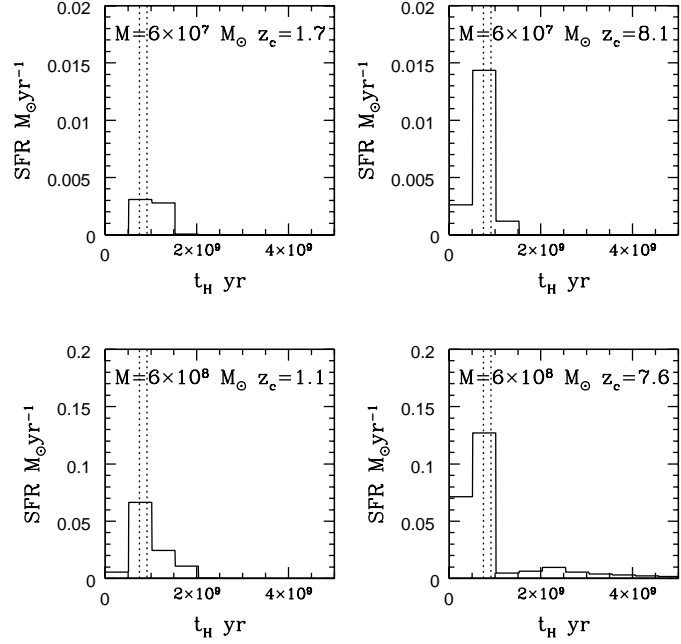}
\caption[dummy]{ Same as Fig. \ref{fig:sfr}, except that $c_*=0.01$.}
\label{fig:vlowsf_sfr}
\end{center}
\end{figure}

\begin{figure}
\begin{center}
\includegraphics[height=10cm]{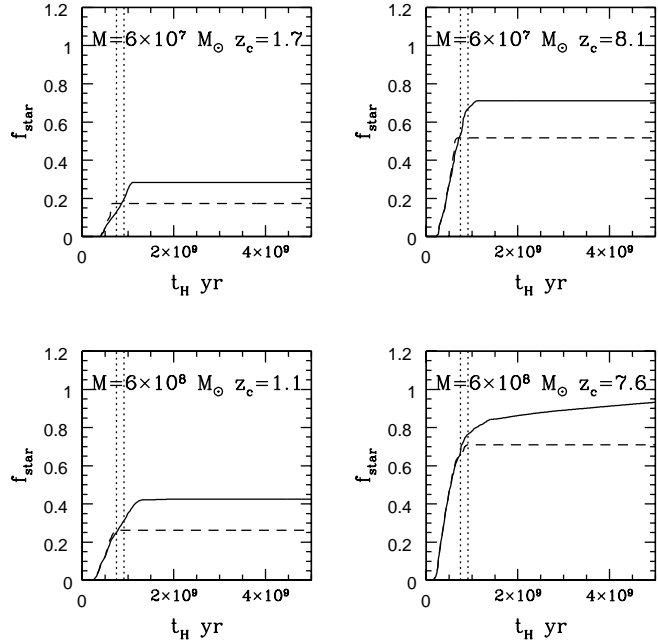}
\caption[dummy]{ The time-variations of stellar fraction $f_{\rm star}$ 
are plotted against the cosmic time for
transfer and optically-thin cases. Solid lines are the results
with radiative transfer, while dashed lines are those under
the assumption of optically-thin. Parameters of four panels are the same
 as that in Fig.\ref{fig:sfr}. 
The vertical short
 dashed lines denote $z=7$ and $z=6$, between which the reionization is almost
 completed in this model.}
\label{fig:f_star_evol}
\end{center}
\end{figure}

\begin{figure}
\begin{center}
\includegraphics[height=10cm]{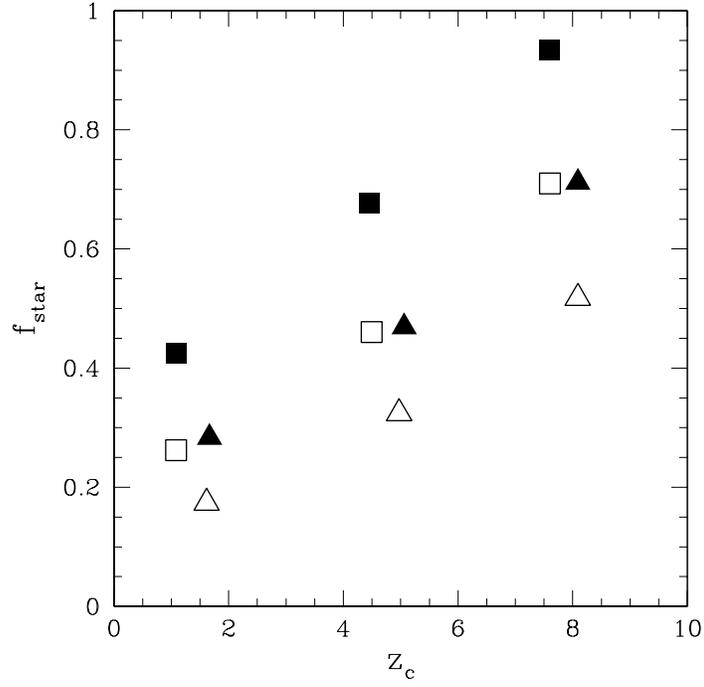}
\caption[dummy]{ The final stellar fraction $f_{\rm star}$ is compared
between radiative transfer and optically-thin cases, assuming
 $c_*=0.1$. Filled symbols represent the results 
of full radiative transfer simulations, and
open symbols denote the results of optically-thin simulations.}
\label{fig:f_star_thin}
\end{center}
\end{figure}

\begin{figure}
\begin{center}
\includegraphics[height=10cm]{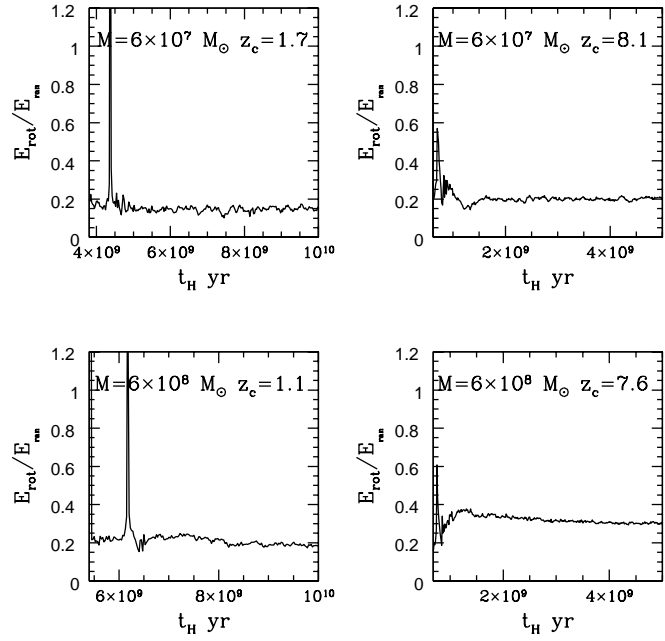}
\caption[dummy]{ The rotation energy-to-random motion energy ratio
($E_{\rm rot}/E_{\rm ran}$) is plotted against the cosmic time. 
Parameters of four panels are the same
 as that in Fig.\ref{fig:sfr}.
Spikes in this figure come from merger of subclumps. 
Since $E_{\rm rot}$ and $E_{\rm ran}$ are measured from the center of
gravity (see the text), violent mergers cause high spikes. 
Especially, in the left panels, the collapse
epoch is relatively late, so that such violent merger phases are conspicuous. }

\label{fig:ratio}
\end{center}
\end{figure}

\begin{figure}
\begin{center}
\includegraphics[height=10cm]{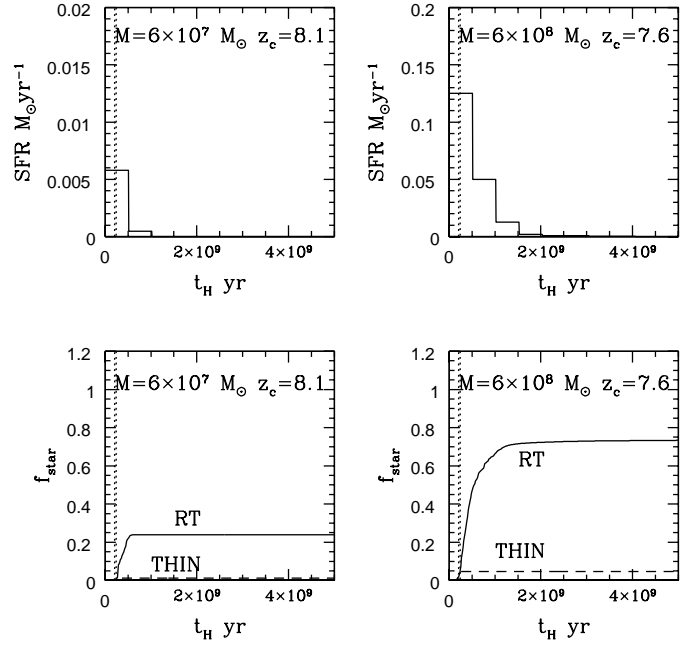}
\caption[dummy]{Star formation histories in the early reionization regime
 are shown (upper two panels), and time evolutions of stellar fraction are
 shown by the solid curves (lower two panels). The parameters are attached
 in each panel. The results of optically-thin simulations are also shown
 in the lower two panels (dashed curve). In these simulations, $c_*=0.1$
 is assumed. The vertical short dashed lines denote $z=17$ and z=16,
between which the reionization is almost
 completed in this early reionization model.}
\label{fig:highz_sfr}
\end{center}
\end{figure}

\begin{figure}
\begin{center}
\includegraphics[height=10cm]{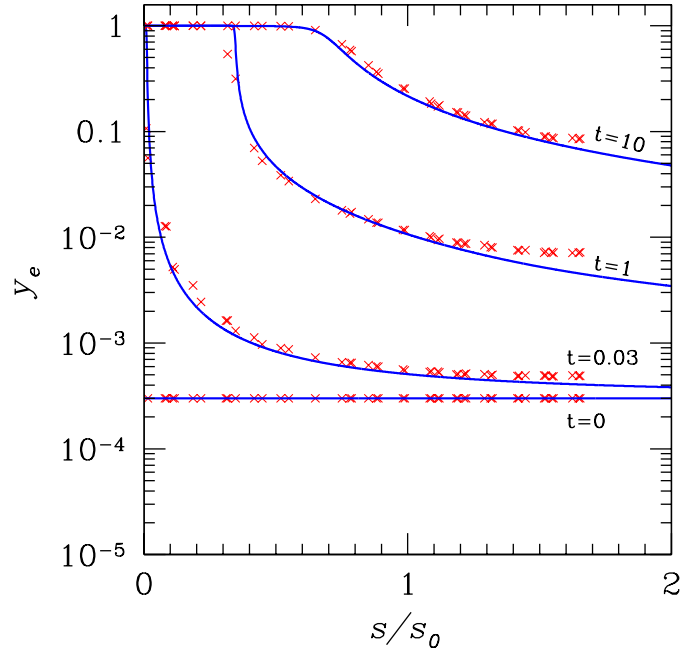}
\caption[dummy]{Propagation of ionization front is shown (see also the text).
Crosses denote the numerical results, while solid lines are analytic solution.}
\label{fig:ion}
\end{center}
\end{figure}

\begin{figure}
\begin{center}
\includegraphics[height=10cm]{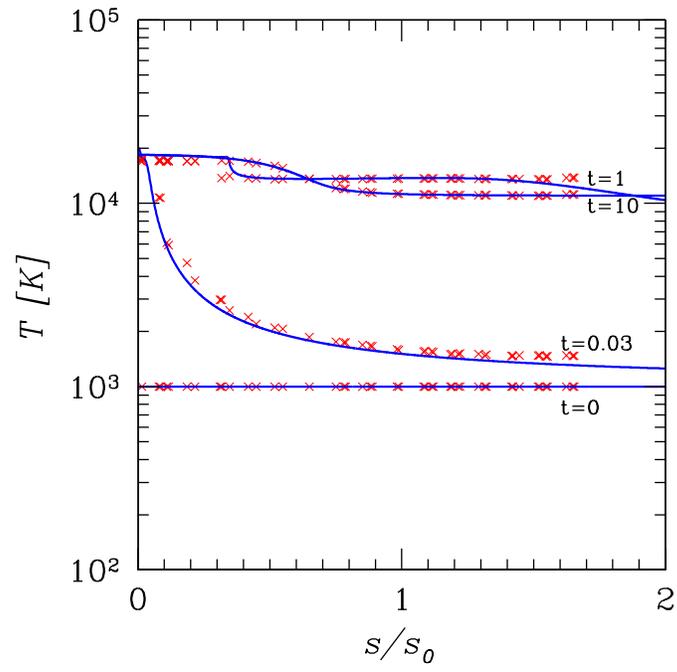}
\caption[dummy]{Same as Fig.\ref{fig:ion}, except that the vertical axis
 denotes gas temperature.}
\label{fig:temperature}
\end{center}
\end{figure}

\end{document}